\documentclass{iau} 
\usepackage{graphicx}

\usepackage{xspace}
\usepackage[numbers]{natbib}

\newcommand{\Al}{$^{26}$Al\xspace}

\newcommand{\Fe}{$^{60}$Fe\xspace}

\newcommand{\Ni}{$^{56}$Ni\xspace}
\newcommand{\Ti}{$^{44}$Ti\xspace}

\newcommand{\Msol}{M\ensuremath{_\odot}\xspace}

\title[Gamma-ray spectroscopy] 
{Gamma-ray line measurements from supernova explosions}

\author[Roland Diehl]   
{Roland Diehl}

\affiliation{Max Planck Institut f\"ur extraterrestrische Physik \\ Giessenbachstr. 1, D-85748 Garching, Germany \\ email: {\tt rod@mpe.mpg.de} \\[\affilskip]
}

\pubyear{2017}
\volume{331}  
\jname{SN 1987A, 30 years later}
\editors{A. Marcowith, G. Dubner, A. Ray, A.Bykov, \& M. Renaud, eds.}
\begin{document}

\maketitle

\begin{abstract}
Gamma ray lines are expected to be emitted as part of the afterglow of supernova explosions, because
radioactive decay of freshly synthesised nuclei occurs. Significant radioactive gamma ray line
emission is expected from \Ni and \Ti decay on time scales of the initial explosion (\Ni, $\tau\sim$days) and the young supernova remnant (\Ti,$\tau\sim$90~years ). Less specific, and rather informative for the supernova population as a whole, are lessons from longer lived isotopes such as \Al and
\Fe. From isotopes of elements heavier than iron group elements, any interesting gamma-ray line emission is too faint to be observable. 
Measurements with space-based gamma-ray telescopes have
obtained interesting gamma ray line emissions from two core collapse events, Cas A and SN1987A, and one thermonuclear event, SN2014J. 
We discuss INTEGRAL data from all above isotopes, including all line and continuum signatures
from these two objects, and the surveys for more supernovae, that have been performed by gamma
ray spectrometry. Our objective here is to illustrate what can be learned from gamma-ray line emission properties about the explosions and their astrophysics.
\keywords{gamma rays: observations, (stars:) supernovae: general, (stars:) supernovae: individual (Cas A, SN1987A, SN2014J), (ISM:) supernova remnants, Galaxy: kinematics and dynamics
}
\end{abstract}

\firstsection 
\section{Introduction}
Supernova explosions have been associated with cosmic matter in hot and dense environments where nuclear reactions can change its composition. 
The light emitted by a supernova is typically found to rise in brightness within few days and then last for months; these characteristics are plausibly connected with the production of radioactive elements in the explosion itself, from such nuclear fusion reactions. The release of this radioactive energy into the expanding explosion thus is a plausible driver of observed supernova light curves\footnote{The explosion blast and even its recombination effect after explosion has ionised the matter, these would only result in brief luminosity lasting typically hours to days at most}.

The \Ni isotope is nature's most-stable n/p-symmetric atomic nucleus, and has a prominent and specific role in the context of supernovae, with its decay first to $^{56}$Co $\tau$=8.8~days, or T$_{1/2}$=6.1~days, and thereafter from $^{56}$Co to stable $^{56}$Fe $\tau$=111.7~days, or T$_{1/2}$=77.4~days. In particular, the near-exponential decline of supernova brightness with time and its similarity to the $^{56}$Co decay time was often taken as supportive -- although in view of the rapidly-expanding supernova and complex radiation-deposit and -transport and -escape physics, this nowadays appears more as a coincidence.

The production of significant amounts of \Ni in all 'normally-bright' supernovae has been a consensus, supported by successful modeling of supernova brightness behaviour based on \Ni as energy source and radiation transport physics alone. Typically, core-collapse supernovae are though to produce on the order of 0.05~\Msol of \Ni, while thermonuclear supernovae produce about ten times as much. 

\begin{figure}  
\begin{center}
 \includegraphics[width=\textwidth]{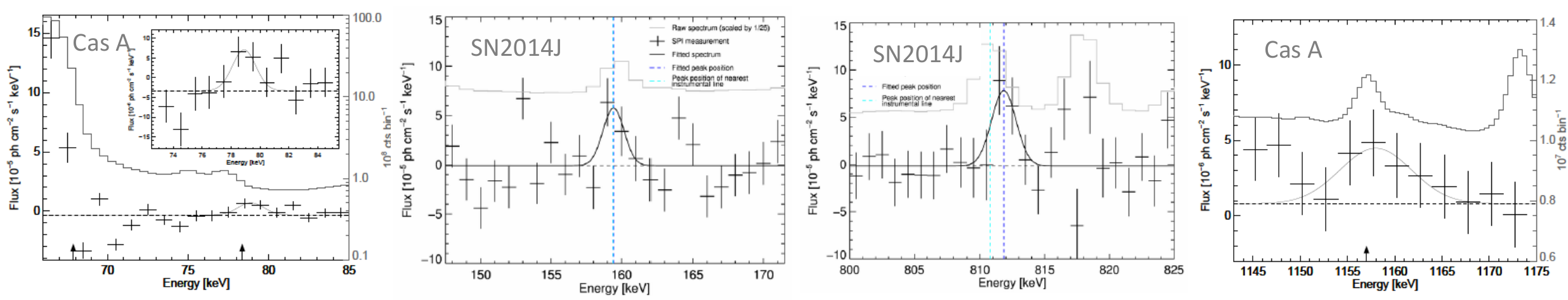} 
 \caption{Examples of gamma-ray line detections from supernovae with the SPI spectrometer on INTEGRAL. Shown are \emph{(from left to right, data points)}: Cas A \Ti line 78 keV \citep{Siegert:2015}, SN2014J \Ni lines 158 keV and 812 keV \citep{Diehl:2014}, and Cas A \Ti line 1157 keV \citep{Siegert:2015},  and raw data \emph{(histograms)}, scaled-down in intensity, to illustrate instrumental-background spectra and their line features.}
   \label{fig_SNline-set}
\end{center}
\end{figure}

It has always been desirable to observe \Ni decay directly though characteristic gamma rays, thus avoiding complex radiation transport as part of the interpretation of supernova brightness. But in the absence of sufficiently sensitive telescopes and/or sufficiently nearby supernovae, this only became a reality rather recently (see Fig.~\ref{fig_SNline-set}), decades after much detail already had been disclosed from analysis of spectra and light curves in other spectral bands, from infrared to optical, UV and X-rays. 

\begin{figure}[t]   
\begin{center}
 \includegraphics[width=0.6\textwidth]{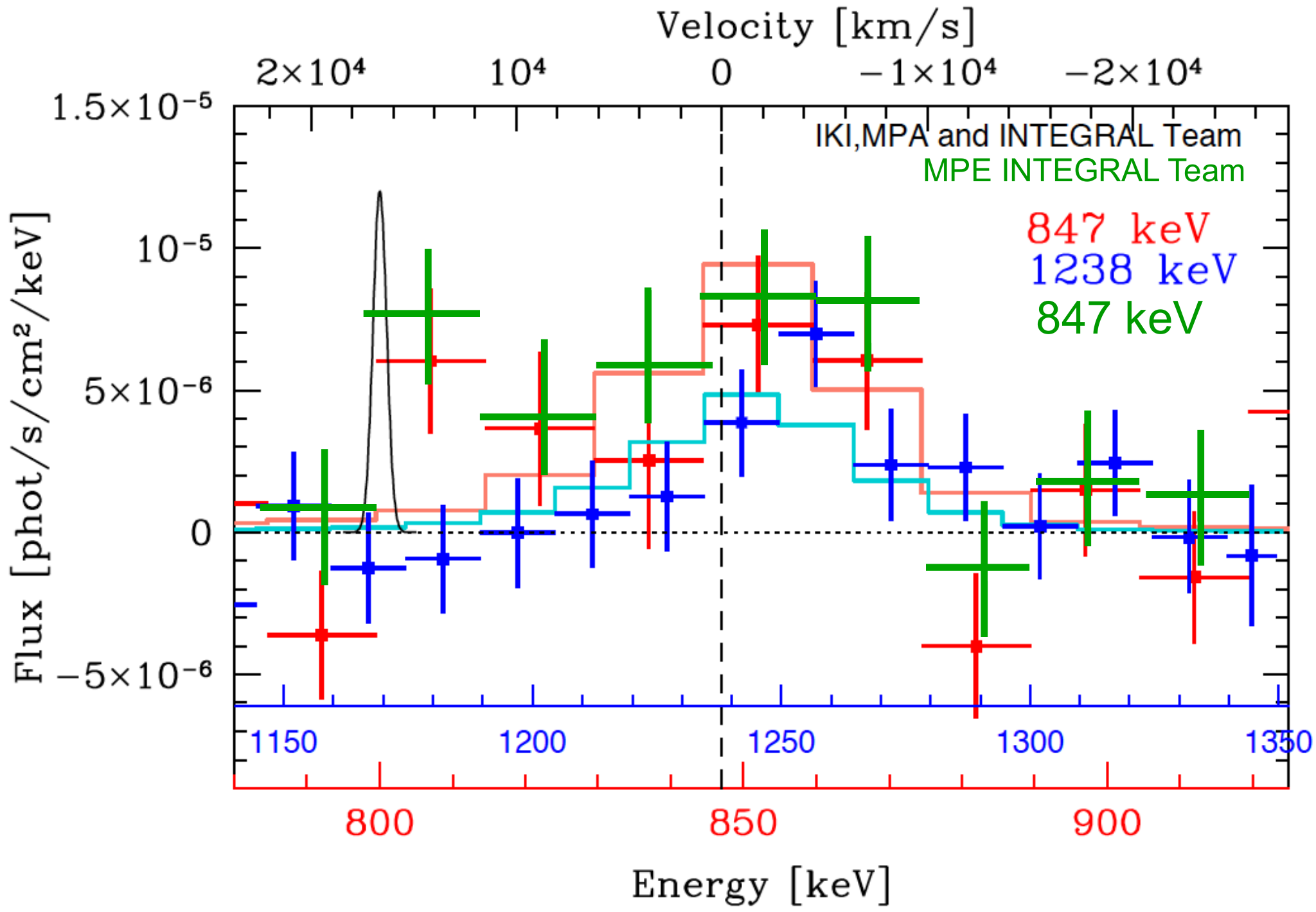} 
 \caption{The $^{56}$Co decay gamma ray lines as observed in SN2014J. Shown are superpositions of the line profiles of the 1238 (blue) and 847 keV lines (red, green) from this decay; the 847 keV analysis with two different analysis methods shows a consistent result in spectral detail. Doppler broadening leads to the observed broad lines, and appears consistently for both lines of $^{56}$Co decay. The Gaussian black line profile shows how a line as narrow as instrumental spectral resolution would appear. }
   \label{fig_SN2014J_lineprofile}
\end{center}
\end{figure}

High resolution gamma ray spectroscopy began from first high-resolution detector balloons and satellites using solid state detectors with Ge as a main ingredient towards resolution $E/\Delta E$ of about 600 in the 1970ies \citep{Mahoney:1982,Leventhal:1978}. The evolution of missions and instrument since then now culminates with SPI on INTEGRAL \citep{Vedrenne:2003,Diehl:2013c}. The competing interests of astronomers for costly space missions make unlikely a successing next-generation mission in coming decades.
Although the explosive nature leads to kinematic broadening of gamma-ray lines of several tens of keV, observing lines with keV-resolution instruments can be meaningful (see below and Fig.~\ref{fig_SN2014J_lineprofile}).

In this paper, we address \Ni gamma rays, and, by analogy, also discuss the astrophysical messages inherent to such type of gamma ray spectroscopy and other line emission from supernovae and their remnants. 

\section{Gamma rays and the physics of supernova explosions}

Thermonuclear supernova explosions are believed to originate from ignition of Carbon burning of degenerate matter inside a white dwarf at densities exceeding 10$^{12}$g~cm$^{-3}$. Nuclear reactions herein approach nuclear statistical equilibrium at least for a fraction of the white dwarf material, thus rearranging nucleons towards nuclei with a maximum of binding energy. The successful empirical description of the nuclear-burning conditions and their evolution in the W7 model standard \citep{Nomoto:1984} predicts a total \Ni mass around 0.5~\Msol \citep[see also][as another standard reference]{Iwamoto:1999}. 
Also many other model variants for thermonuclear supernovae have predicted \Ni production within factors of a few of this \citep[see reviews in][]{Ropke:2011aa,Hillebrandt:2013aa}, and \Ni radioactive energy dominates the energy input as the supernova obtains it brightness maximum, typically 20 days after the onset of the explosion \citep[see discussion of 3D models and other radioactivities in][]{Seitenzahl:2013aa}. 

The main diagnostic power of gamma-ray observations lies in the brightness evolution of the characteristic gamma-ray lines from the \Ni decay chain, which includes lines from the first decay to $^{56}$Co as well as to the second and final product $^{56}$Fe, and the 511~keV line from positrons produced in the $\beta^+$-decay. 
The expanding supernova is initially dense enough to absorb all such gamma rays, thus converting energy to lower-energy radiation and internal energy of the expanding gas. As the supernova expands, more and more of the gamma rays will escape, thus becoming observable as their energy deposit in the supernova fades away. 
The brightness evolution thus reflects both the distribution of \Ni within the supernova and the absorbing overlying material. Measurements of the gamma ray brightness evolution therefore can \emph{calibrate} the measurements of re-radiated radioactive energy, as they constrain the energy source. 
Moreover, the spatial distribution of \Ni may become observable rather directly, e.g. distinguishing inside-out explosions with \Ni enclosed in a large mass of white dwarf material from explosion scenarios where nuclear burning occurs partly or predominantly in outer regions of the exploding object.

Also core collapse of a massive star leads to high-density matter suitable for nuclear rearrangements and even nuclear statistical equilibrium \citep{Nomoto:2006a}. But in this case, the infalling and exploding gas dynamics is less constrained, thus may evolve over the time of collapse and onset of the explosion, and vary for different progenitor masses. 
As a result, the amount of \Ni expected from such explosions is much less and also more uncertain and variable, as equilibrium conditions only occur in a relatively small region of the available parameter space \citep{Thielemann:2011,Magkotsios:2010}. 
Moreover, the larger amount of  overlying material in the outer shells above the collapsing iron core results in modest expectations of observable \Ni gamma ray emission, compared to the case of thermonuclear explosions. But another radioactive nucleus with a half life of about 60 years becomes interesting in this case, depositing radioactive energy mainly through its positrons from $\beta^+$-decay in the late supernova. At such late times the supernova is fully transparent to \Ti decay gamma rays, and these therefore allow to observe kinematics and spatial distribution of \Ti as a representative product of nuclear processing deep in the original supernova \citep{Thielemann:2011,Magkotsios:2010}. 

Other line emission in the nuclear energy range may be expected at somewhat later times, as particle acceleration in young supernova remnant reaches a peak efficiency. This is expected within the first few hundred years of the remnant evolution, as ambient gas is swept up by the still rapidly-expanding blast wave. Accelerated nuclei will result in collisionally-excited nuclei which de-excite with emission of characteristic nuclear lines \citep{Ramaty:1979}.

Late emission of nuclear lines from long-lived radioactivities of \Al and \Fe from supernova ejecta provides another diagnostic about the cosmic importance of supernovae: The spatial distribution of ejecta at late times, as it mixes with ambient interstellar gas, determines the time scales of recycling of newly-produced nuclei from one generation of supernovae towards a next generation of stars that forms from enriched cold interstellar gas.      

\section{Core collapse events: SN1987 and Cas A}

\begin{figure}   
\begin{center}
 \includegraphics[width=0.9\textwidth]{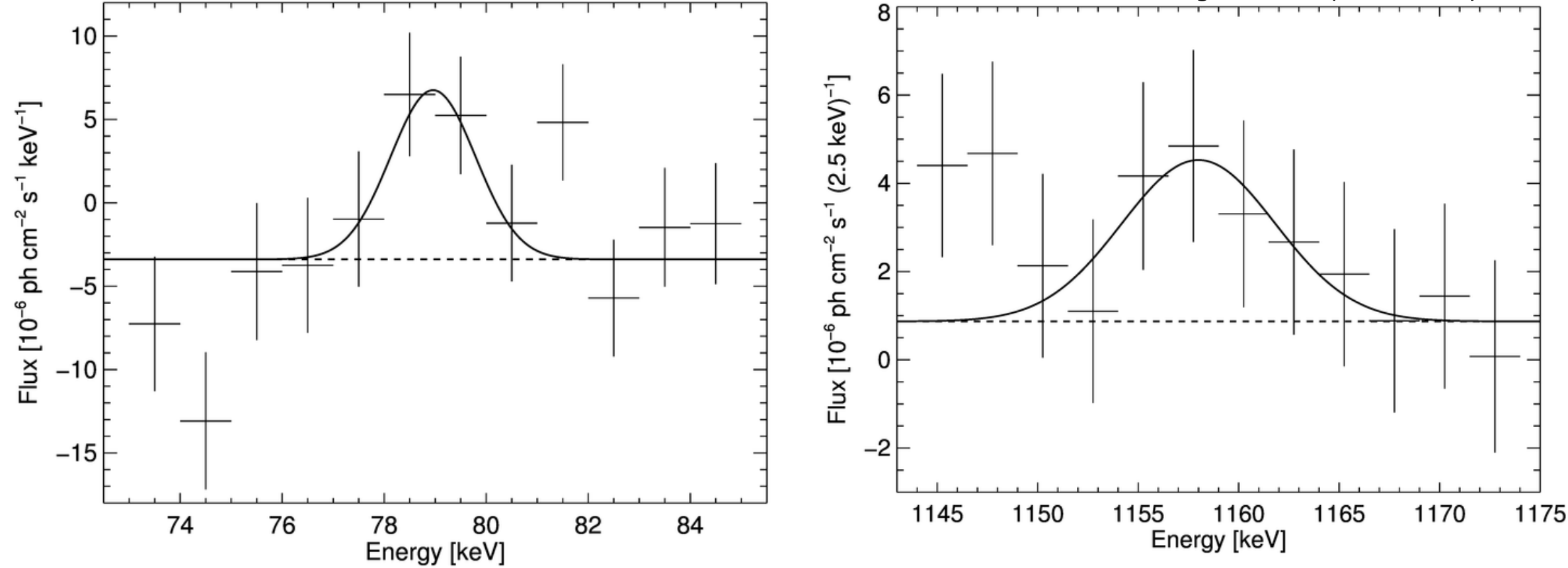} 
 \caption{The $^{44}$Ti decay gamma ray lines as observed in Cas A with SPI on INTEGRAL \citep{Siegert:2015}.  Doppler broadening leads to the observed broad lines. }
   \label{fig_CasA_lineprofiles}
\end{center}
\end{figure}  

SN1987A was the first supernova where gamma ray lines were reported:  Measured with the Solar Maximum Missions gamma-ray spectrometer \citep{Matz:1988}, this nicely confirmed the \Ni origin of supernova light. This result moreover provided strong indications for explosion asymmetry, because the lines from $^{56}$Co decay appeared much earlier than expected  (see above). 
The GRIS balloon mission with a Ge detector provided the first measurement of the 847 keV $^{56}$Co line profile \citep{Tueller:1990}, indicating the redshift that seemed plausible from the early appearance. Later, INTEGRAL reported detection of \Ti emission from SN1987A \citep{Grebenev:2012}, which was also measured with better spectroscopic quality by NuSTAR more recently \citep{Boggs:2015}. 
Thus, SN1987A is not only the best-observed core collapse supernova that shapes what we believe today about this type of explosion \citep{McCray:2016}, it also is a prominent example of the success of multi-wavelenth science.

The second supernova of type core collapse, which has been studied in many astronomical windows, is Cas A \citep{Vink:2004}. This supernova exploded already more than 350 years ago, so that \Ni amounts can only inferred indirectly here, from light echoes. The young Cas A supernova remnant shows several remarkable properties, such as fast-moving knots, and an apparent overturn of the original layering of material of its progenitor star.
  
 \Ti emission from Cas A has  been detected first with COMPTEL. A diversity of efforts followed to confirm and improve upon this exciting possibility to measure \Ti radioactiivity from a supernova. Results were reported from OSSE on CGRO \citep{The:1996}, RXTE \citep{Rothschild:1999}, Beppo-Sax \citep{Vink:2001}, and both INTEGRAL instruments IBIS and SPI \citep{Renaud:2006a,Martin:2009a,Siegert:2015}, culminating with NuSTAR's observations where NuSTAR's X-ray mirror allowed to spatially resolve this emission \citep{Grefenstette:2014} . 
This first image of \Ti gamma-ray line emission \citep{Grefenstette:2014} provides a key lesson on supernova nucleosynthesis and morphology, with interesting deviations from a spherical explosion \citep{Grefenstette:2017} (see also Grefenstette et al., Janka et al., Fryer et al., and Wongwhatanarath, these proceedings).
INTEGRAL/SPI data are being accumulated to refine spectroscopic results, as only SPI is capable to measure both the hard X-ray and gamma ray lines from the \Ti decay chain \citep{Siegert:2015}. Preliminary latest analysis (Weinberger et al., in preparation) provides interesting hints of a composite line structure, compatible with and supporting the latest spatially-resolved spectroscpic results from NuSTAR \citep{Grefenstette:2017}. More data will be collected in the coming years of the continuing INTEGRAL mission.

\begin{figure}   
\begin{center}
 \includegraphics[width=0.6\textwidth]{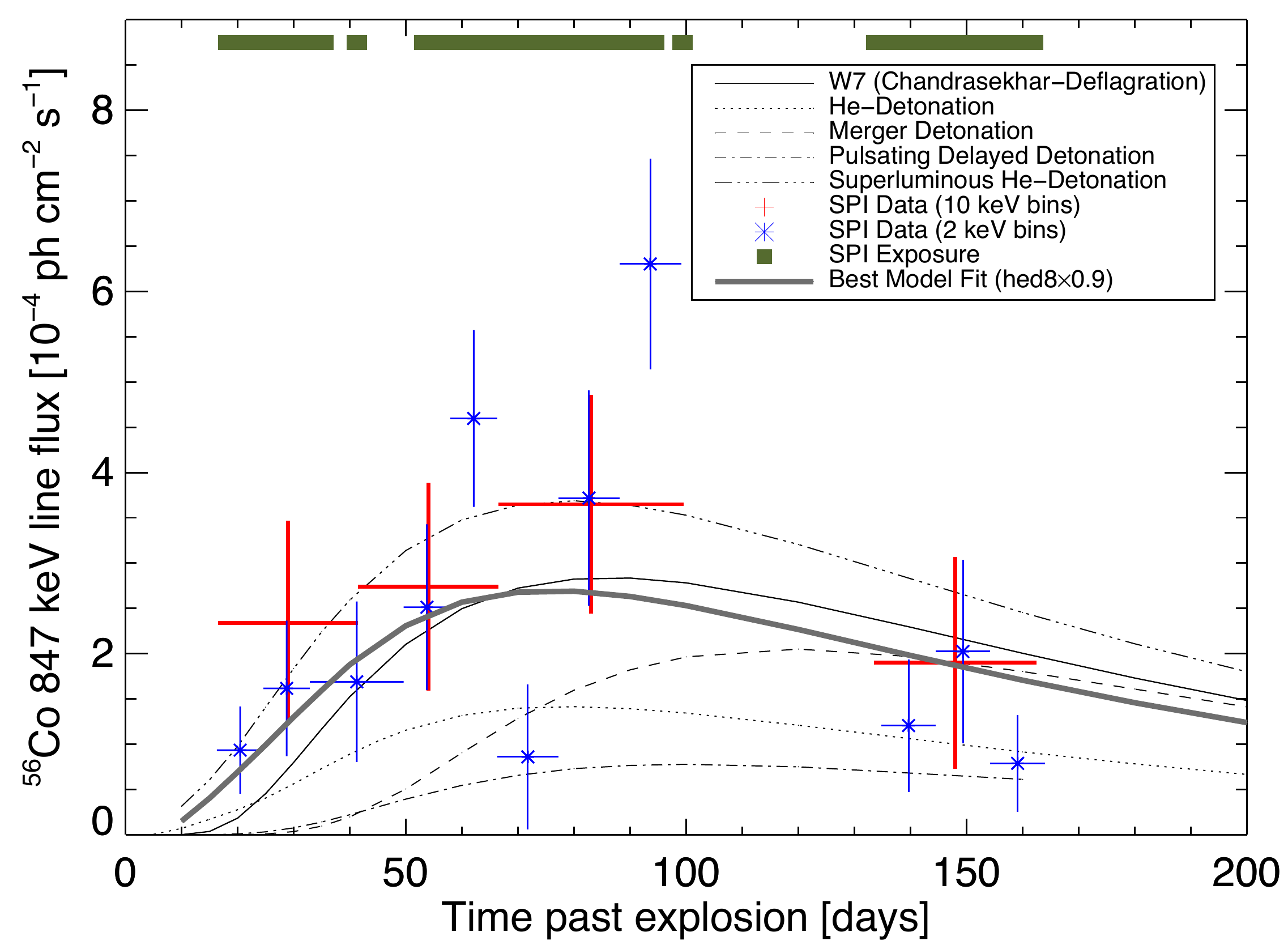} 
 \caption{The $^{56}$Co decay gamma ray line brightness evolution of SN2014J, as seen with SPI on INTEGRAL \citep{Diehl:2015}. }
   \label{fig_SN2014J_lightcurve}
\end{center}
\end{figure}  

\section{Thermonuclear supernovae: SN2014J}
The first gamma-ray line results from a thermonuclear supernova had been long awaited, and were realised with SN2014J and INTEGRAL's instruments \citep{Churazov:2014,Diehl:2015}. This provides an independent measurement of the \Ni mass ejected \citep{Diehl:2015}. 
As a surprise, early line emission from the shortlived \Ni decay could be seen as well \citep{Diehl:2014}. This indication of \Ni near the surface, but also the irregular rise of the $^{56}$Co emission line brightness (Fig.~\ref{fig_SN2014J_lightcurve}) \citep{Diehl:2015}, all provide strong indications that 1-dimensional models of a spherical explosion are too restrictive a model, and do not adequate represent reality. 

\section{Supernova explosions throughout the Galaxy}
\subsection{Environments of supernova explosions}
Supernovae of type Ia may occur also in regions remote from the birth sites of their progenitors, while core collapse supernovae result from massive stars that are born in clusters and evolve within millions of years, so exploding near their formation sites. Therefore, \Ti emitting young supernova remnants were the objective of searches with gamma-ray surveys of the Compton \citep{Dupraz:1997,Iyudin:1999a} and INTEGRAL \citep{Renaud:2006,Tsygankov:2016} Observatories. 
The lessons from both surveys are \citep{The:2006,Dufour:2013} that typical core-collapse supernovae are not ejecting much of \Ti, while the ones with indications of strong deviations from spherical symmetry sometimes do. Interestingly, also from Tycho a hint of \Ti emission has been reported \citep{Troja:2014,Wang:2014}, while this event is commonly associated with a thermonuclear explosion.  

The \Al map \citep{Diehl:1995b} provides yet another survey of nucleosynthesis ejecta from supernovae, as its origins had been associated with massive star origins predominantly \citep{Prantzos:1996a}. 
Active regions as showing up in \Al have been studied in more detail. Interesting for the topic of this paper is the Orion-Eridanus region \citep{Diehl:2003}, where several massive star groups are located on the near side of the Orion molecular clouds, and a large interstellar cavity, the Eridanus bubble, extends from those towards the Sun. In theoretical simulations and their comparisons with X- and gamma-ray emission of the region, it seems that supernova explosions drive the evolution and continuum emission characteristics of such superbubbles \citep{Krause:2014a}.

\subsection{How supernovae shape properties of the Galaxy}
\begin{figure}   
\begin{center}
 \includegraphics[width=0.2\textwidth]{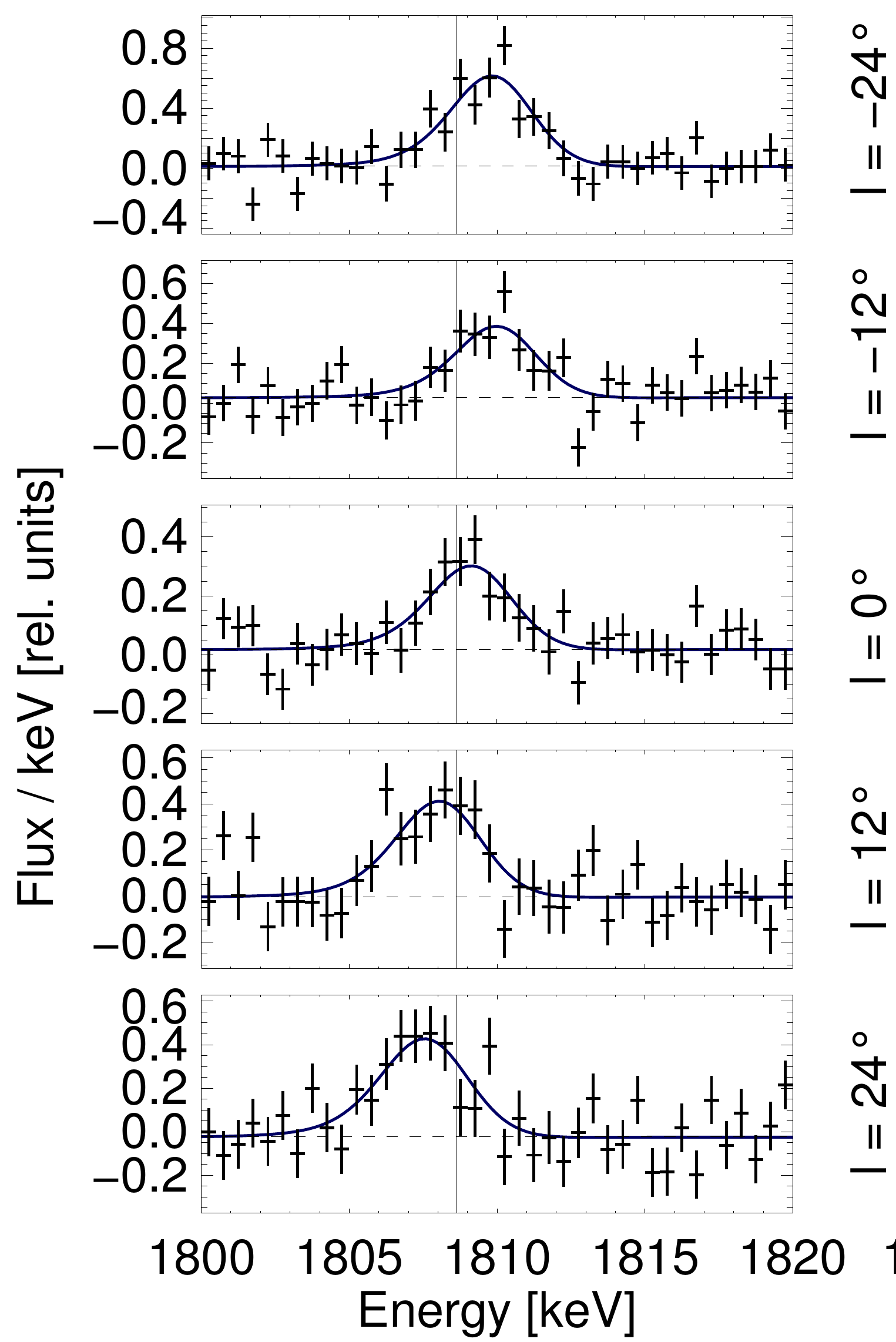} 
  \includegraphics[width=0.44\textwidth]{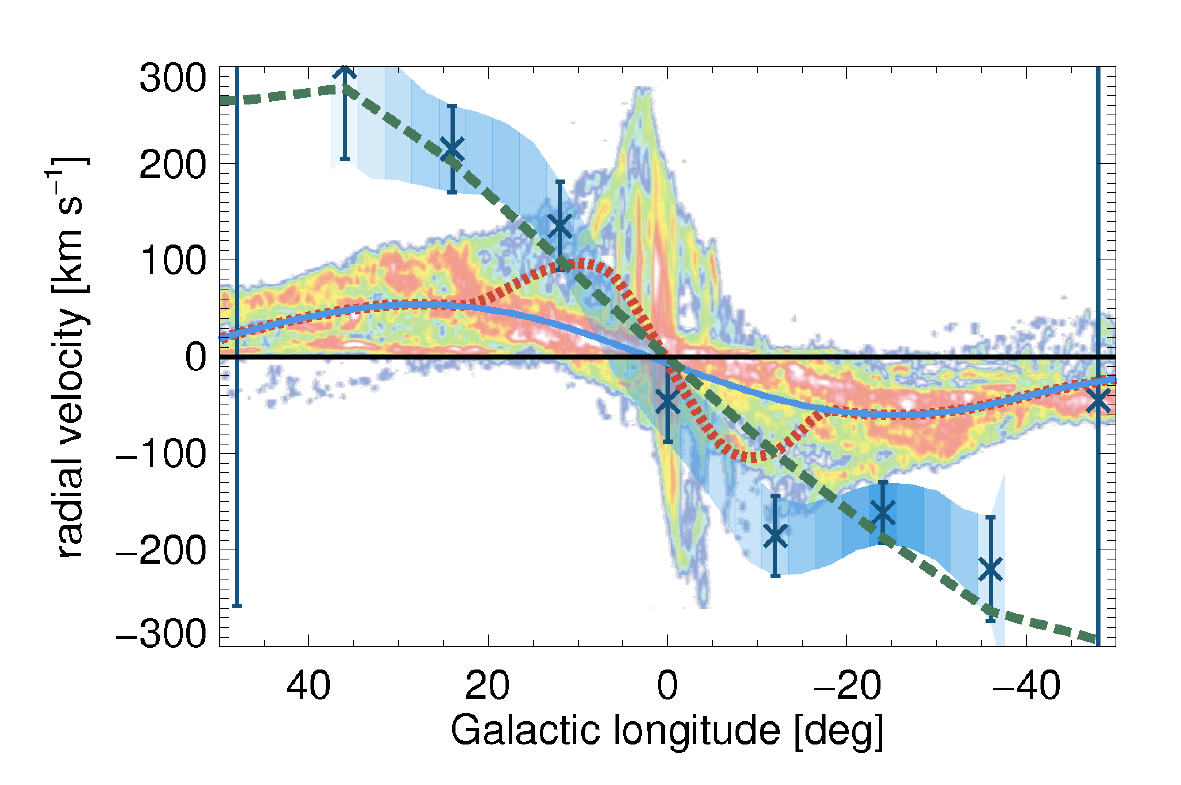} 
   \includegraphics[width=0.32\textwidth]{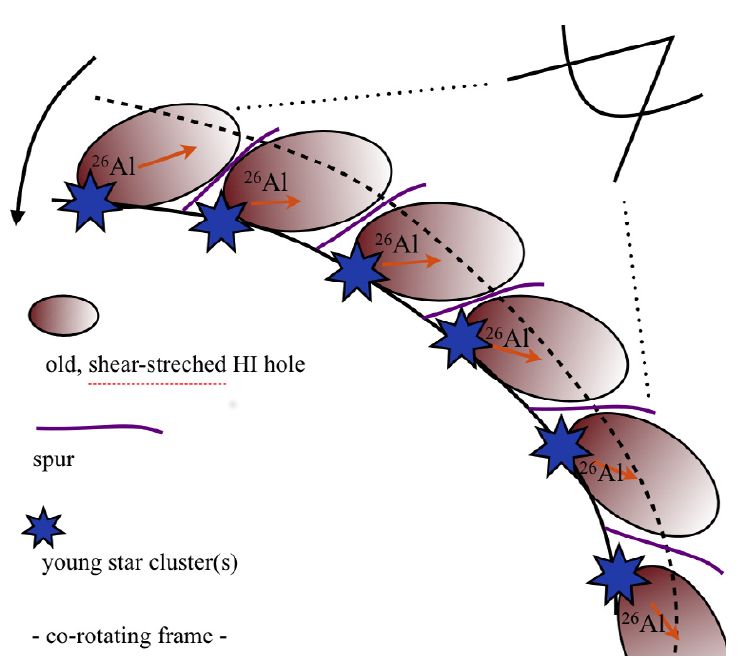} 
 \caption{The $^{26}$Al line lessons on superbubbles as recycling regions. }
   \label{fig_26Al-superbubbles}
\end{center}
\end{figure}

The superbubble creation and evolution that has become apparent in the Orion-Eridanus example seems to even operate galaxy-wide: The \Al line shape measurements with SPI/INTEGRAL have revealed systematic Doppler shifts from large-scale galactic rotation (Fig.~\ref{fig_26Al-superbubbles} left) \citep{Kretschmer:2013}, that exceeds expectations from other tracers of galactic rotation by about 200~km~s$^{-1}$ (Fig.~\ref{fig_26Al-superbubbles} center). This suggests that \Al ejection from massive-star groups systematically occurs into asymmetric superbubbles, that are formed as those massive-star groups have moved to the leading edge of spiral arms during the evolution time of the massive stars before their \Al ejection (Fig.~\ref{fig_26Al-superbubbles} right) \citep{Krause:2015}. 
This means that SN ejecta occur as a mixture of cold cavity walls and hot bubble interiors. As superbubbles break up in the later phase of their evolution, ejecta thus may escape in champagne flows or winds, and thus take considerable time to cool and condense as part of sites of new star formation:
Recycling of SN ejecta appears to be a more complex process, that may not be captured in a single recycling time scale, and is not \emph{instantaneous}, strictly speaking.

\bibliographystyle{aa}


\newcommand{\actaa}{Acta Astron. }%
\newcommand{\araa}{Ann.Rev.Astron.\&Astroph. }%
\newcommand{\apj}{Astroph.J. }%
\newcommand{\apjl}{Astroph.J.Lett. }%
\newcommand{\apjs}{Astroph.J.Supp. }%
\newcommand{\ao}{Appl.~Opt. }%
\newcommand{\apss}{Astroph.J.\&Sp.Sci. }%
\newcommand{\aap}{Astron.\&Astroph. }%
\newcommand{\aapr}{Astron.\&Astroph.~Rev. }%
\newcommand{\aaps}{Astron.\&Astroph.~Suppl. }%
\newcommand{\aj}{Astron.Journ. }%
\newcommand{\azh}{AZh }%
\newcommand{\memras}{MmRAS }%
\newcommand{\mnras}{Mon.Not.Royal~Astr.~Soc. }%
\newcommand{\na}{New Astron. }%
\newcommand{\nar}{New Astron. Rev. }%
\newcommand{\pra}{Phys.~Rev.~A }%
\newcommand{\prb}{Phys.~Rev.~B }%
\newcommand{\prc}{Phys.~Rev.~C }%
\newcommand{\prd}{Phys.~Rev.~D }%
\newcommand{\pre}{Phys.~Rev.~E }%
\newcommand{\prl}{Phys.~Rev.~Lett. }%
\newcommand{\pasa}{PASA }%
\newcommand{\pasp}{Proc.Astr.Soc.Pac. }%
\newcommand{\pasj}{Proc.Astr.Soc.Jap. }%
\newcommand{\rpp}{Rep.Prog.Phys. }%
\newcommand{\skytel}{Sky\&Tel. }%
\newcommand{\solphys}{Sol.~Phys. }%
\newcommand{\sovast}{Soviet~Ast. }%
\newcommand{\ssr}{Space~Sci.~Rev. }%
\newcommand{\nat}{Nature }%
\newcommand{\iaucirc}{IAU~Circ. }%
\newcommand{\aplett}{Astrophys.~Lett. }%
\newcommand{\apspr}{Astrophys.~Space~Phys.~Res. }%
\newcommand{\nphysa}{Nucl.~Phys.~A }%
\newcommand{\physrep}{Phys.~Rep. }%
\newcommand{\procspie}{Proc.~SPIE }%
         




\end{document}